\documentstyle[12pt,amsfonts]{article}
\parskip=\medskipamount

\newcommand {\cC}{{\cal C}}
\newcommand {\cD}{{\cal D}}

\newcommand {\cF}{{\cal F}}

\newcommand {\cK}{{\cal K}}
\newcommand {\cL}{{\cal L}}

\newcommand {\cN}{{\cal N}}

\newcommand {\cR}{{\cal R}}

\newcommand {\cT}{{\cal T}}

\newcommand {\cY}{{\cal Y}}

\newcommand{\bA}{{\bf A}}
\newcommand{\bB}{{\bf B}}

\def\a{\alpha}
\def \bi{\bibitem}

\def\b{\beta}
\def\c{\chi}
\def\d{\delta}
\def\e{\epsilon}
\def\f{\phi}
\def\g{\gamma}
\def\G{\Gamma}

\def\j{\psi}
\def\k{\kappa}
\def\l{\lambda}
\def\m{\mu}

\def\o{\omega}

\def\q{\theta}
\def\r{\rho}
\def\s{\sigma}

\def\F{\Phi}
\def\J{\Psi}
\def\L{\Lambda}
\def\O{\Omega}

\def\S{\Sigma}
\def\U{\Upsilon}

\newcommand{\ad}{{\dot{\alpha}}}                           
\newcommand{\bd}{{\dot{\beta}}}                            
\newcommand{\ve}{\varepsilon}                            
\newcommand{\cDB}{{\bar \cD}}                            
\newcommand{\gd}{{\dot{\gamma}}}

\newcommand{\pa}{\partial}                           
\newcommand{\hf}{\frac12}

\newcommand{\vf}{\varphi}

\newcommand{\be}{\begin{equation}}
\newcommand{\ee}{\end{equation}}
\newcommand{\bea}{\begin{eqnarray}}
\newcommand{\eea}{\end{eqnarray}}
\newcommand{\non}{\nonumber}

\def \intss{\int\!\!{\rm d}^8z}

\def \ERc{\frac{E^{-1}}{R}}
\def \ERac{\frac{E^{-1}}{\bar R}}
\def \Dsqc{(\cD^2 - 4 {\bar R})}
\def \Dsqac{(\cDB^2 - 4 R)}
\newcommand{\la}{\langle}
\newcommand{\ra}{\rangle}

\begin{document}

\begin{titlepage}

\begin{flushright}
hep-th/0501172\\
January, 2005
\end{flushright}
\vspace{5mm}

\begin{center}
{\Large\bf  
On the component structure of \\
$\cN=1$ supersymmetric nonlinear electrodynamics
}
\end{center}
\vspace{3mm}

\begin{center}

{\large Sergei M. Kuzenko and Shane A. McCarthy}
\vspace{2mm}

\footnotesize{
{\it School of Physics M013, The University of Western Australia,\\
35 Stirling Highway, Crawley W.A. 6009, Australia}} \\
{\tt kuzenko@cyllene.uwa.edu.au}~,
{\tt shane@physics.uwa.edu.au}
\vspace{2mm}

\end{center}
\vspace{5mm}

\begin{abstract}
\baselineskip=14pt
We analyze the component structure of models
for 4D $\cN=1$ supersymmetric nonlinear 
electrodynamics that enjoy invariance 
under continuous duality rotations.  
The $\cN=1$ supersymmetric Born-Infeld action 
is a member of this family. Such dynamical systems 
have a more complicated structure, 
especially in the presence of supergravity, 
as compared with well-studied effective 
supersymmetric theories containing at  most 
two derivatives (including nonlinear K\"ahler sigma-models). 
As a result, when deriving their  canonically normalized 
component  actions, it becomes 
impractical and cumbersome to follow the traditional 
approach of (i) reducing to components; and 
then (ii) applying a field-dependent Weyl and local 
chiral transformation. It proves  to be more efficient 
to follow the Kugo-Uehara scheme  which 
consists of (i) extending the superfield theory 
to a super-Weyl invariant system; and then 
(ii) applying a plain component reduction along 
with imposing a suitable super-Weyl gauge condition.
Here we implement this scheme to derive the bosonic
action of self-dual supersymmetric electrodynamics 
coupled to the dilaton-axion chiral multiplet
and a K\"ahler sigma-model.
In the fermionic sector, the action contains higher 
derivative terms.
In the globally supersymmetric case, 
a nonlinear field redefinition 
is explicitly constructed which eliminates 
all the higher derivative terms and brings the 
fermionic action to a one-parameter deformation of 
the Akulov-Volkov action for the Goldstino.  
The Akulov-Volkov action emerges, in particular, 
in the case of the $\cN=1$ supersymmetric 
Born-Infeld action.
\end{abstract}
\vfill
\end{titlepage}

\newpage
\setcounter{page}{1}
\renewcommand{\thefootnote}{\arabic{footnote}}
\setcounter{footnote}{0}

\section{Introduction}
\setcounter{equation}{0}
The Born-Infeld theory \cite{BI} is a particular representative 
in the family of models for nonlinear electrodynamics 
which are 
grouped together through
a single classification principle of self-duality, that is
invariance under continuous electromagnetic 
duality rotations \cite{GZ1,GR,GZ}.
The requirement of self-duality is equivalent 
to the fact that the Lagrangian $L(F)$ is a solution 
to the (non-supersymmetric) self-duality equation 
 \cite{GR,GZ}
$$
\tilde{F}^{ab} \,F_{ab} + \tilde{G}^{ab} \,G_{ab} = 0~, 
\qquad
\tilde{G}^{ab}(F) \equiv  \hf \ve^{abcd}\,G_{cd}(F) = 
2\, \frac{\pa L(F) }{\pa F_{ab}}~, 
$$
with $\tilde{F}$ the Hodge-dual of $F$.
What makes the Born-Infeld model unique is, 
in particular, its appearance as a low-energy  
effective action in string theory \cite{FT,L}.
It is worth mentioning that a
 general theory of (nonlinear) self-duality 
in four and higher space-time  dimensions 
for non-supersymmetric theories was developed
in \cite{list}.

The concept of self-dual nonlinear electrodynamics 
\cite{GR,GZ} was extended to 4D $\cN=1, ~2$ 
globally supersymmetric theories \cite{KT,KT2}.
Such a marriage of nonlinear electromagnetic 
self-duality with supersymmetry has turned out  
to be quite robust, since the families of actions 
obtained include all the known models 
for partial breaking of supersymmetry based 
on the use of a vector Goldstone multiplet.
In particular, the $\cN=1$ supersymmetric Born-Infeld
action \cite{CF}, which  is a Goldstone multiplet 
action for partial supersymmetry 
breakdown $\cN=2 \to \cN=1$ \cite{BG,RT},
appears, at the same time,  to be a solution 
to the $\cN=1$  self-duality equation 
\cite{KT,KT2}.  Furthermore, the model for partial 
breaking of supersymmetry $\cN=4 \to \cN=2$
developed in \cite{BIK} 
proves to be a unique solution to 
the $\cN=2$ self-duality equation 
possessing a nonlinearly realized central charge 
symmetry \cite{KT2}. 

Self-dual supersymmetric electrodynamics can 
naturally be coupled to superfield supergravity  
\cite{KM}, 
using either the old minimal \cite{WZ-old,old}
or the new minimal \cite{new}
formulations  of $\cN=1$ supergravity
(see textbooks \cite{GGRS,WB,BK}  for reviews
on superfield supergravity).  
As demonstrated in \cite{KM},
such dynamical systems possess quite 
remarkable properties including (i) duality-invariance 
of the supercurrent; (ii) self-duality 
under Legendre transformation. 
These properties are a natural 
generalization of similar properties 
in the non-supersymmetric case \cite{GR,GZ} and 
 in the globally supersymmetric case \cite{KT,KT2}.
An unexpected feature of self-dual locally supersymmetric
systems is that they couple not only to the dilaton-axion 
chiral multiplet (that transforms under duality rotations), 
but also to those nonlinear K\"ahler sigma-models
which are inert under duality rotations. 

While the considerations in  \cite{KT,KT2,KM}
were given mainly in terms of superspace and 
superfields,  here we would like to subject to scrutiny   
the component structure of self-dual 
supersymmetric systems. 
This turns out  to involve two 
rather  nontrivial  aspects.

${}$For general supergravity-matter systems
with {\it at  most two derivatives} at the component level
\cite{Cremmer1,Cremmer2}, 
the traditional approach (reviewed in \cite{WB})
of obtaining canonically normalized 
component  actions consists of two steps:
(i)  a plain reduction from superfields to components;
(ii) the application of a field-dependent Weyl and local 
chiral transformation (accompanied by a gravitino shift).  
Now, as we turn to nonlinear 
supersymmetric electrodynamics, 
a generic term in the component action 
may involve {\it any number of derivatives} -- 
already the purely electromagnetic part of the Lagrangian, 
$L(F)$, is a nonlinear function of the field strength.
${}$For such supergravity-matter systems, 
the traditional approach can be argued 
to become impractical and cumbersome 
(as regards the component tensor calculus employed 
in  \cite{Cremmer1,Cremmer2}, it  has never been 
extended, to the best of our knowledge,  
to the case of the supersymmetric theories 
we are going to  study below, 
therefore the superspace approach 
is the only formalism at our disposal). 
There exist two alternatives \cite{Kugo,BGG} to the
traditional approach of component reduction \cite{WB}
that were originally developed for the systems scrutinized 
in \cite{Cremmer1,Cremmer2} or slightly more 
generals ones, but remain equally powerful 
in a more general setting. We prefer to 
follow the Kugo-Uehara
approach \cite{Kugo} that conceptually 
originates in \cite{Kaku}
and is quite natural  in the framework of 
the Siegel-Gates formulation of superfield 
supergravity \cite{Siegel}.
The idea is to follow the pattern of  the Weyl 
invariant extension of Einstein gravity, 
\bea
S[g] = {1 \over 2 \k^2} \int {\rm d}^4x 
\sqrt{-g} \,R \quad \longrightarrow \quad
S[g, \vf ] = {3 \over \k^2} \int {\rm d}^4x 
 \sqrt{-g} \,
\Big\{ g^{mn} \, \pa_m \vf \pa_n \vf 
+ \frac{1}{6} R \, \vf^2 \Big\}~,
\non
\eea
and extend any supergravity-matter system 
to a super-Weyl invariant system 
(in the Howe-Tucker sense  \cite{HT}) by introducing 
a compensating  covariantly chiral scalar superfield
$\S$ (in addition to the supergravity chiral 
compensator \cite{Siegel}).
When reducing to components, 
canonically  normalized component actions 
are obtained simply by imposing a suitable 
super-Weyl gauge condition 
to effectively eliminate $\S$.

In the present paper, 
we implement the component reduction 
scheme of \cite{Kugo} to derive the bosonic
action of self-dual supersymmetric nonlinear 
electrodynamics 
coupled to the dilaton-axion chiral multiplet
and a K\"ahler sigma-model.
As concerns the fermionic sector, 
the situation is highly nontrivial even 
in the globally supersymmetric case. 
The point is that the fermionic action contains higher 
derivative terms that seem to be removable 
in the presence of supergravity fields. 
In the globally supersymmetric case, 
we explicitly construct  a nonlinear field redefinition 
which eliminates 
all the higher derivative terms and brings the 
fermionic action to a one-parameter deformation of 
the Akulov-Volkov action for the Goldstino \cite{VA,AV}.  
The Akulov-Volkov action emerges, in particular, 
in the case of the $\cN=1$ supersymmetric 
Born-Infeld action.

This paper is organized as follows.
In section 2 we review, following \cite{BK},
the procedure of reducing locally supersymmetric 
actions from superfields to components.
In section 3 we  then spell out the 
Kugo-Uehara scheme \cite{Kugo} 
on the example of 
a nonlinear K\"ahler sigma-model 
coupled to supergravity.
Section 4 is devoted to the derivation of the 
bosonic action of self-dual supersymmetric 
nonlinear  electrodynamics 
coupled to the dilaton-axion chiral multiplet
and a K\"ahler sigma-model.
Different aspects of the fermionic dynamics
in the globally supersymmetric case 
are analyzed in sections 5 and 6. 
A discussion of the results obtained 
and future perspectives 
is given in section 7. Some nuances
of the Akulov-Volkov (AV) action are presented in
appendix A. In particular, we demonstrate 
that all the terms of eighth order in the AV action 
completely cancel. Finally, appendix B 
is devoted to an alternative realization 
 of old minimal supergravity.

\section{From superfield supergravity
\setcounter{equation}{0}
to components}

Here we recall salient points of the old minimal 
and the new minimal formulations of $\cN=1$ 
supergravity (see \cite{GGRS,WB,BK} for more
details), and also review, following \cite{BK},
the procedure of reducing locally supersymmetric 
actions from superfields to components.
\subsection{Old minimal supergravity}
We follow the notation\footnote{
In particular, $z^M = (x^m, \q^\m , {\bar \q}_{\dot \m})$ are the
coordinates of $\cN=1$ curved superspace,
${\rm d}^8z = {\rm d}^4 x\, {\rm d}^2 \q \,{\rm d}^2 {\bar \q}$ is
the full flat superspace measure, and
${\rm d}^6z = {\rm d}^4 x\, {\rm d}^2 \q$ is the measure in the
chiral subspace.}
and $\cN=1$ supergravity conventions of \cite{BK}.
Unless otherwise stated we work with the old
minimal formulation of $\cN=1$ supergravity.
The superspace geometry is described 
by covariant derivatives
\bea
\cD_A &=& (\cD_a , \cD_\a ,\cDB^\ad ) = E_A + \O_A~, \non \\
E_A &=& E_A{}^M (z) \pa_M  ~, \qquad
\O_A = \hf\,\O_A{}^{bc} (z) M_{bc}
= \O_A{}^{\b \g} (z) M_{\b \g}
+\O_A{}^{\bd \gd} (z) {\bar M}_{\bd \gd} ~,
\eea
with $E_A^{~M} $ the vielbein,
$\O_A $ the Lorentz connection
and $M_{bc} \Leftrightarrow ( M_{\b\g}, {\bar M}_{\bd \gd})$
the Lorentz generators.
The covariant derivatives obey the following algebra:
\bea
\label{algebra}
&& {} \qquad \{ \cD_\a , {\bar \cD}_\ad \} = -2{\rm i} \cD_{\a \ad} ~, 
\non \\
\{\cD_\a, \cD_\b \} &=& -4{\bar R} M_{\a \b}~, \qquad
\{ {\bar \cD}_\ad, {\bar \cD}_\bd \} = - 4R {\bar M}_{\ad \bd}~,  \\
\left[ { \bar \cD}_{\ad} , \cD_{ \b \bd } \right]
     & = & -{\rm i}{\ve}_{\ad \bd}
\Big(R\,\cD_\b + G_\b{}^{\dot{\g}}  \cDB_{\dot{\g}}
-(\cDB^\gd G_\b{}^{\dot{\d}})
{\bar M}_{\gd \dot{\d}}
+2W_\b{}^{\g \d}
M_{\g \d} \Big)
- {\rm i} (\cD_\b R)  {\bar M}_{\ad \bd}~,  \non  \\
\left[ \cD_{\a} , \cD_{ \b \bd } \right]
     & = & \phantom{-}{\rm i}{\ve}_{\a \b}
\Big({\bar R}\,\cDB_\bd + G^\g{}_\bd \cD_\g
- (\cD^\g G^\d{}_\bd)  M_{\g \d}
+2{\bar W}_\bd{}^{\gd \dot{\d}}
{\bar M}_{\gd \dot{\d} }  \Big)
+ {\rm i} (\cDB_\bd {\bar R})  M_{\a \b}~,  \non
\eea
where the tensors $R$, $G_a = {\bar G}_a$ and
$W_{\a \b \g} = W_{(\a \b\g)}$ satisfy the Bianchi identities
\be
\cDB_\ad R= \cDB_\ad W_{\a \b \g} = 0~, \quad
\cDB^\gd G_{\a \gd} = \cD_\a R~, \quad
\cD^\g W_{\a \b \g} = {\rm i} \,\cD_{(\a }{}^\gd G_{\b) \gd}~.
\ee
Modulo purely gauge degrees of freedom, 
all geometric objects --
the vielbein and the connection -- can be expressed
\cite{Siegel}
in terms of three unconstrained superfields
(known as the prepotentials of 
old minimal supergravity):
gravitational superfield $H^m = {\bar H}^m$,
chiral compensator $\vf$  (${\bar E}_\ad \vf =0$)
and its conjugate $\bar \vf$.
The old minimal supergravity action is
\be
\label{omsg}
S_{\rm SG,old} = - 3
\intss\,E^{-1}~, \qquad \quad
E = {\rm Ber} (E_A{}^{M})~,
\ee
with the gravitational coupling constant 
being set equal to one.

\subsection{New minimal supergravity}
We will also deal with the new minimal 
formulation of supergravity.
This can be treated (see \cite{BK} for a review)
as a super-Weyl invariant dynamical system
that couples old minimal supergravity  
to the improved tensor multiplet \cite{deWR}
described by  a {\it real} covariantly
linear scalar superfield ${\Bbb L}$ \cite{Siegel2},
\be
\Dsqac\, {\Bbb L} =
(\cD^2 - 4 {\bar R})\, {\Bbb L} = 0~.
\label{linear}
\ee

Super-Weyl transformations, originally introduced in \cite{HT},
are simply local rescalings of the chiral compensator in 
old minimal supergravity \cite{Siegel}
(see also \cite{GGRS,BK}),
\be
\vf  ~\to~ {\rm e}^{\s} \, \vf~,
\ee
with $\s(z) $ an arbitrary  covariantly chiral scalar parameter,
$\cDB_\ad \s=0$. In terms of the covariant derivatives,
the transformation\footnote{Under
(\ref{superweyl}), the full superspace measure changes as
${\rm d}^8 z\, E^{-1} \to
{\rm d}^8 z\, E^{-1} \,\exp (\s + \bar \s ) $,
while the {\it chiral superspace measure} transforms as
${\rm d}^8 z\, E^{-1} / R \to
{\rm d}^8 z\, (E^{-1} /R ) \, \exp (3\s )$.} is
\be
\cD_\a ~\to~ {\rm e}^{ \s/2 - {\bar \s} } \Big(
\cD_\a - (\cD^\b \s) \, M_{\a \b} \Big) ~, \qquad
\cDB_\ad ~\to~ {\rm e}^{ {\bar \s}/2 - \s } \Big(
\cDB_\ad -  (\cDB^\bd {\bar \s}) {\bar M}_{\bd\ad} \Big)~,
\label{superweyl}
\ee

Since
\be
(\cD^2 - 4 {\bar R}) ~ \to ~{\rm e}^{-2 \bar \s} \,
(\cD^2 - 4 {\bar R})\,{\rm e}^{ \s}
\label{quadr}
\ee
when acting on a scalar superfield, it is clear
that the super-Weyl transformation law of $\Bbb L$ is
uniquely fixed  to be
\be
{\Bbb L} ~\to ~ {\rm e}^{-\s - \bar \s} \, {\Bbb L}~.
\label{s-weyl-L}
\ee
The new minimal supergravity action is
\be\label{nmsg}
S_{\rm SG,new} = 3 \intss\, E^{-1}\,
{\Bbb L}\, {\rm ln} {\Bbb L}~.
\ee

Any system of matter superfields $\J$ coupled to
new minimal supergravity can be treated as
a super-Weyl invariant coupling
of old minimal supergravity to the matter
superfields $\J$ and $\Bbb L$ 
(see \cite{BK} for a review).

Old minimal and new minimal supergravities are
dual to each other. One can show this by considering
the ``first-order'' action
\be\label{aux1}
S = 3 \int\!\!{\rm d}^8z \,E^{-1}\,(U\, {\Bbb L}
- {\rm e}^{U})~,
\ee
where $U(z)$ is an arbitrary real scalar superfield. For
(\ref{aux1}) to be super-Weyl invariant, $U$ must
transform under super-Weyl transformations as
\be\label{UWeylTransform}
U ~\to~ U -\s - \bar \s~.
\label{s-weyl-U}
\ee
Solving for the equation of motion of $U$, we regain
the new minimal supergravity action (\ref{nmsg}). On
the other hand, the solution to the equation of motion
for ${\Bbb L}$ requires that $U$ be the sum of a
covariantly chiral scalar superfield and its conjugate,
\be
\label{Usolution}
U = {\rm ln} \S + {\rm ln} {\bar \S}~,\quad\qquad
\cDB_\ad \S = 0
\ee
The action (\ref{aux1}) then becomes
\be
\tilde{S}_{\rm SG,old}  
= - 3 \intss\,E^{-1}\,{\bar \S}\,\S~,
\label{omsg2}
\ee
where $\S$ has, 
in accordance with (\ref{s-weyl-U}),
the following super-Weyl
transformation
\be
\label{SWeylTransform}
\S ~\to~ {\rm e}^{-\s}\, \S~.
\label{s-weyl-S}
\ee
An alternative realization of this 
dynamical system is given in Appendix B.

The action (\ref{omsg2}) 
is the super-Weyl invariant extension of the
old minimal supergravity action (\ref{omsg}).
The latter 
may be recovered by using the super-Weyl gauge
freedom to impose the gauge condition $\S=1$. 
In what follows, we prefer to use (\ref{omsg2}). 
Any dynamical system of matter superfields $\J$ 
coupled to old minimal supergravity can be 
promoted to a a super-Weyl invariant system 
if the chiral compensator $\vf$ is replaced by 
the super-Weyl invariant combination 
\be
\vf~ \to ~\vf \, \S~.
\label{convertion}
\ee
This is equivalent to applying 
the super-Weyl transformation 
(\ref{superweyl}) with $\s=\S$
(which may be accompanied by 
a super-Weyl transformation of the matter 
superfields).

\subsection{Components in old minimal supergravity}
The old minimal  supergravity multiplet 
$\{{e_{a}}^{m}, {\J_{a}}^{\b},
{\bar \J}_{a \bd} ,  A_{a},
B, {\bar B}\}$ comprises  the (inverse) vierbein
${e_{a}}^{m}$, the gravitino 
${\bf \J}_a = ( {\J_{a}}^{\b},
{\bar \J}_{a \bd} )$, 
and the auxiliary fields\footnote{These auxiliary 
fields are denoted as $\bA_a$, $ \bB$ and
${\bar \bB}$ in \cite{BK}.} 
 $A_a$, $ B$ and ${\bar B}$.
Within the framework of superfield 
supergravity, these component fields
naturally appear  in a Wess-Zumino gauge
\cite{WZ}  (see  \cite{GGRS,WB,BK} for reviews).  
Here we use the Wess-Zumino gauge chosen in 
\cite{BK}.

We define superfields'  component fields
by space projection and covariant differentiation. 
${}$For a superfield $V(z)$, the former 
 is the zeroth
order term in the power series expansion in $\q$ and
${\bar \q}$
\be
V\arrowvert = V(x,\q = 0, {\bar \q} = 0)~.
\ee
The space projection of the vector covariant derivatives are
\be
\cD_{a}\arrowvert = \nabla\!_{a}
- \frac{1}{3}\ve_{abcd}\, {A}^{d} M^{bc}
+ \frac{1}{2} {\J_{a}}^{\b} \, \cD_{\b}\arrowvert
+ \frac{1}{2} {\bar \J}_{a\bd} \, {\bar \cD}^{\bd}
\arrowvert ~,
\ee
where we have introduced the spacetime 
covariant derivatives, 
$\nabla\!_{a} = e_{a} + \frac{1}{2} \o_{abc} M^{bc}$, 
with 
$\o_{abc}= \o_{abc}(e,\J) $ the connection 
and $e_{a}={e_{a}}^{m}\partial_{m}$. 
The explicit expressions for the projections 
$\cD_\a|$ and ${\bar \cD}^\ad|$ can be 
found in \cite{BK}.
The spacetime  covariant derivatives 
obey the following algebra
\be
\label{eq:covariant derivative algebra}
\left[\nabla\!_{a}, \nabla\!_{b}\right] = {\cT_{ab}}^{c}\,\nabla\!_{c}
+ \frac{1}{2}\,\cR_{abcd} M^{cd}~,
\ee
where $\cR_{abcd}$ is the curvature tensor and $\cT_{abc}$ 
is the torsion. The torsion is related to the gravitino by
\be
\cT_{abc} = -\frac{\rm i}{2}(\J_{a}\s_{c}{\bar \J}_{b}
- \J_{b}\s_{c}{\bar \J}_{a})~.
\ee
Additionally we can write the connection in terms of the 
supergravity fields as
\be
\o_{abc}  
= \o_{abc}(e) - \frac{1}{2}(\cT_{bca}+\cT_{acb}-\cT_{abc})
~, \qquad 
\o_{abc}(e) = \frac{1}{2}(\cC_{bca}+\cC_{acb}-\cC_{abc}) ~,
\ee
where $\cC_{abc}$ are the anholonomy coefficients,
\be
\left[e_{a},e_{b}\right] =
{\cC_{ab}}^{c}e_{c}~,
\quad\qquad {\cC_{ab}}^{c} =
\left((e_{a}{e_{b}}^{m})
- (e_{b}{e_{a}}^{m})\right) {e_{m}}^{c}~.
\ee

The supergravity auxiliary fields occur as 
follows
\be
R\arrowvert = \frac{1}{3}{B}~,\quad\qquad
G_{a}\arrowvert = \frac{4}{3}{ A}_{a}~.
\ee
One also has 
\bea
\cD_{\a}R\arrowvert &=& 
-\frac{2}{3} (\s^{bc}\J_{bc})_{\a}
-\frac{2{\rm i}}{3} {A}^{b}\J_{b \a}
+\frac{\rm i}{3} {\bar {B}} (\s^{b}{\bar \J}_{b})_{\a}~,\non\\
{\bar \cD}_{(\ad}{G^{\b}}_{\bd)}\arrowvert &=&
-2{\J_{\ad\bd,}}^{\b}+\frac{\rm i}{3}{\bar { B}}
{{\bar \J}^{\b}}\,_{(\ad,\bd)}
-2{\rm i}({\tilde \s}^{ab})_{\ad\bd}{\J_{a}}^{\b}A_{b}
+\frac{2{\rm i}}{3}{\J_{\a(\ad,}}^{\a}A_{\bd)}{}^\b~,\non\\
W_{\a\b\g}\arrowvert &=& \J_{(\a\b,\g)}
-{\rm i}(\s_{ab})_{(\a\b}
{\J^a}_{\g)}A^{b}~,
\eea
and
\bea
{\bar \cD}^{2}{\bar R}\arrowvert &=& \frac{2}{3} \left(\cR
+ \frac{\rm i}{2} \ve^{abcd}\, \cR_{abcd}\right)
+\frac{16}{9}A^{a}A_{a}
+\frac{4}{9}\e^{abcd}\cT_{abc}A_{d}
-\frac{8{\rm i}}{3}(\nabla\!_{a}A^{a})
+\frac{8{\rm i}}{9}{\cT_{ab}}^{b}A^{a}
\non\\
&&+\frac{8}{9} {B}{\bar {B}}
+ \frac{4}{9} { B} (\J_{a}\s^{ab}\J_{b})
+{\rm i} {\bar \cD}_{\ad}{\bar R}\arrowvert
({\tilde \s}^{a}\J_{a})^{\ad}
+ \frac{2{\rm i}}{3}\J^{\a\ad,\b}
\cD_{(\a}G_{\b)\ad}\arrowvert~,
\eea
where
\bea
&&\qquad\qquad
\J_{ab}{}^\g = \nabla\!_{a}\J_{b}{}^\g - \nabla\!_{b}\J_{a}{}^\g
- {\cT_{ab}}^{c}\J_{c}{}^\g~,\non\\
&&\J_{\a\b,}{}^\g=
\frac{1}{2}(\s^{ab})_{\a\b}\J_{ab}{}^\g
~,\quad\qquad
\J_{\ad\bd,}{}^\g=
-\frac{1}{2}({\tilde \s}^{ab})_{\ad\bd}\J_{ab}{}^\g~.
\eea
With these objects and the covariant derivative algebra,
(\ref{algebra}) the method to obtain the component action is 
as follows.

Since
\be
\intss\,E^{-1}\,\cL = -\frac14 \intss\, \ERc\, \Dsqac \cL
~,
\label{chiralrule}
\ee
modulo a total derivative,
it is sufficient to work with chiral 
actions involving 
a chiral scalar Lagrangian  $\cL_{\rm c}$, 
${\bar \cD}_{\ad}\cL_{\rm c} = 0$. 
Such a chiral action generates the following 
component action \cite{BK}
\bea
\label{reductionrule}
 \int\!\!{\rm d}^8z\,\frac{E^{-1}}{R}\cL_{\rm c}
&=& \int\!\!{\rm d}^4x\,e^{-1}\Big\{-\frac{1}{4}
\cD^2 \cL_{\rm c}\arrowvert 
- \frac{\rm i}{2} ({\bar \J}^{b} 
\tilde{\s}_{b})^{\a}\, \cD_{\a}\cL_{\rm c}\arrowvert
+ (B + {\bar \J}^{a}{\tilde \s}_{ab}{\bar \J}^{b}) \,
\cL_{\rm c}\arrowvert
\Big\}~, 
\non  \\
e &=& \det (e_a{}^m )~.
\eea

The component action for old minimal supergravity, 
(\ref{omsg}) is
\be
\label{compsugra}
S_{\rm SG,old} =\int\!\!{\rm d}^4x\,e^{-1}\Bigg\{\frac{1}{2}\cR
+\frac{4}{3}{A}^{a}{A}_{a} - \frac{1}{3}{\bar { B}} B
+ \frac{1}{4}\ve^{abcd} ({\bar \J}_{a} {\tilde \s}_{b} \J_{cd}
- \J_{a} \s_{b} {\bar \J}_{cd})\Bigg\}~,
\ee
see \cite{BK} for more details.

\section{K\"ahler sigma-models in  supergravity}
\setcounter{equation}{0}
To illustrate the Kugo-Uehara approach to component 
reduction \cite{Kugo}, we  consider  
a nonlinear K\"ahler sigma-model coupled
to supergravity. 

\subsection{Superfield formulations}
K\"ahler sigma-models are most easily 
described within the framework of new
minimal supergravity
(see, e.g. \cite{BK} for a review). Given a
K\"ahler manifold parametrized by $n$ complex coordinates
$\f^{i}$ and their conjugates ${\bar \f}^{\underline i}$, with
$K\!(\f, \bar \f )$ the K\"ahler potential, the corresponding
supergravity-matter action is
\be
\label{sigma}
S= 3 \intss\, E^{-1}\,
{\Bbb L}\, {\rm ln} {\Bbb L} + \intss\, E^{-1}\,
{\Bbb L}\,K\!(\f, \bar \f )~.
\ee
The dynamical variables $\f$ are covariantly chiral scalar
superfields, $\cDB_\ad \f=0 $, being inert with respect to
the super-Weyl transformations.
The action is obviously super-Weyl invariant. 
Due to (\ref{chiralrule}), it is also
invariant under the K\"ahler transformations 
\be
\label{kahler}
K\!(\f, \bar \f) \to K\!(\f, \bar \f) + \l(\f) + \bar \l(\bar \f)~,
\ee
with $\l(\f)$ an arbitrary holomorphic function. 

To reformulate the dynamics
within the framework of old
minimal supergravity, 
let us introduce the auxiliary action
\be
\label{aux2}
S = 3 \intss\, E^{-1}\,
\left(U\, {\Bbb L} - \U\right)~,
\ee
where
\be
\label{upsilon}
\U = \exp\!\Big(U - \frac{1}{3} K\!(\f, \bar \f) \Big) ~,
\ee
and $U$ is an arbitrary real scalar superfield. For the
action (\ref{aux2}) to be super-Weyl invariant $U$ must
transform by the law (\ref{UWeylTransform}) under 
transformations (\ref{superweyl}). 
To preserve K\"ahler invariance (\ref{kahler}),   
the  K\"ahler
transformation of $U$ should be 
\be
U ~\to~ U + \frac{1}{3}\left(\l(\f) + \bar \l(\bar \f)\right)~.
\ee
Solving  the equation of motion for $U$, we regain
the supergravity-matter action (\ref{sigma}). On
the other hand, the solution to the equation of motion
for ${\Bbb L}$ requires that $U$ be the sum of a
covariantly chiral scalar superfield and its conjugate,
as given by (\ref{Usolution}). The K\"ahler sigma-model
then reads
\be
\label{kahlermodel}
S_{\rm Kahler} = -3  \int\!\!{\rm d}^8z\, E^{-1}\,
{\bar \S}\, \S \, 
{\rm exp}\!\left(-\frac 13 K\!(\f,{\bar \f})\right)
=  -3\int\!\!{\rm d}^8z\, E^{-1}\, \tilde{\U}~,
\ee
where $\U$ is as defined in (\ref{upsilon}) but with 
$U$ given by the solution (\ref{Usolution})
\be
{\tilde \U} = \S{\bar \S} \,
{\rm exp}\!\left(-\frac{1}{3} K\!(\f,{\bar \f})\right)~.
\label{tilde-u}
\ee
Super-Weyl
transformations of $\S$ are given by
(\ref{SWeylTransform}), whereas under the K\"ahler
transformations (\ref{kahler}) we have
\be
\S ~\to~ {\rm e}^{\l(\f)/3}\,\S~.
\ee

\subsection{Component action of K\"ahler sigma-model}
To determine the component
structure of (\ref{kahlermodel}), 
the  approach of \cite{Kugo} requires 
that a particular super-Weyl gauge choice
be made such that the Einstein and Rarita-Schwinger
terms in the component action come out in canonical
form.

We define\footnote{With such a definition,
both $\c^{i}$ and $F^{i}$
transform as tangent vectors 
under arbitrary holomorphic
reparametrizations,
$Y^{i} \to f^{i}(Y)$,
of the K\"ahler manifold with K\"ahler potential
$K(Y,{\bar Y})$.} the component fields of the
chiral scalar superfields $\f^{i}$ by
\bea
\label{scalarcompdef}
\f^{i}\arrowvert = Y^{i}~,\qquad
\cD_\a \f^{i}\arrowvert = \c^{i}_{\a}~, \qquad
-\frac{1}{4}\cD^2 \f^{i}\arrowvert = F^{i}
&+& \frac{1}{4}\,{\G^{i}}_{jk} \, \c^{j}\c^{k}~,
\eea
where we have introduced the Christoffel symbols
${\G^{i}}_{jk}$ of the K\"ahler manifold defined by
K\"ahler potential $K (Y,{\bar Y})$. The metric of the
K\"ahler manifold is
\be
g_{i{\underline i}} =g_{{\underline i}i} =
\frac{\partial^{2}K (Y,{\bar Y})}{
\partial Y^{i} \partial {\bar Y}^{\underline i}}
\equiv K_{i{\underline i}}~,
\ee
where the subscript $i$ (${\underline i}$) on $K$
denotes differentiation with respect to $Y^{i}$
(${\bar Y}^{\underline i}$). Similarly, we can write
expressions for the Christoffel symbols and the
curvature on the K\"ahler manifold
\bea
{\G^{i}}_{jk} = g^{i{\underline i}} K_{jk{\underline i}}~,
\quad&~&\quad 
{\G^{\underline i}}_{{\underline j}{\underline k}}
= g^{{\underline i}i} K_{i{\underline j}{\underline k}}~,\non\\
{\rm R}_{ij{\underline i}{\underline j}} =
K_{ij{\underline i}{\underline j}} 
&-& g^{k{\underline k}}
K_{ij{\underline k}} K_{k{\underline i}{\underline j}}
~,
\eea
where the matrix elements 
$g^{i{\underline i}} = g^{{\underline i}i}$
correspond to  the inverse  K\"ahler metric, 
$g^{i{\underline j}} \,g_{{\underline j}k}
=\d^i{}_k$.

Applying the reduction formula (\ref{reductionrule}) to the
K\"ahler sigma model (\ref{kahlermodel}) we obtain
\bea
\label{compexpansion}
S_{\rm Kahler} = -3\int\!\!{\rm d}^4x\,e^{-1}
\Bigg\{&&\!\!\!\left(-\frac{1}{4}\cD^2 R\arrowvert 
- \frac{\rm i}{2} ({\bar \J}^{a}{\tilde \s}_{a})^{\a}\cD_{\a}
R\arrowvert
+ ({B} + {\bar \J}^{a}{\tilde \s}_{ab}{\bar \J}^{b})
R\arrowvert
\right) \tilde{\U} \arrowvert \non\\
&&-\frac{1}{2}\cD^{\a}R\arrowvert\cD_{\a}
\tilde{\U}\arrowvert
-\frac{1}{4}R\arrowvert
\cD^2{\tilde \U}\arrowvert
-\frac{\rm i}{2}({\bar \J}^{a}{\tilde \s}_{a})^{\a}
R\arrowvert\cD_{\a} \tilde{\U} \arrowvert\\
&&-\frac{1}{16}\cD^{2}{\bar \cD}^{2}
\tilde{\U} \arrowvert
+\frac{\rm i}{8}({\bar \J}^{a}{\tilde \s}_{a})^{\a}\cD_{\a}{\bar \cD}^{2}
\tilde{\U} 
\arrowvert
-\frac{1}{4}({ B} + {\bar \J}^{a}{\tilde \s}_{ab}{\bar \J}^{b})
{\bar \cD}^{2} \tilde{\U} 
\arrowvert
\Bigg\} ~.\non
\eea
The first line of (\ref{compexpansion}) reduces to the
supergravity action (\ref{compsugra}) if we make a
super-Weyl gauge choice such that 
$\tilde{ \U} \arrowvert=1$. This can
be done by setting
\be
\S\arrowvert = {\rm e}^{K (Y,\bar Y)/6}~.
\ee
We have now eliminated the need to perform
a Weyl rescaling on the component action. Further
specification of the components of $\S$ can be used to
remove the need for the chiral rotation (and gravitino shift).
We accomplish this with the following choices
\bea
\cD_\a \S\arrowvert &=& \frac{1}{3}\,\c^{i}_{\a}
K_{i}\,{\rm e}^{K(Y,\bar Y)/6}~,\\
-\frac{1}{4}\cD^2 \S\arrowvert &=&
\left(\frac{1}{3} F^{i} K_{i}
- \frac{1}{12}\,\c^{i}\c^{j} (K_{ij} - {\G^{k}}_{ij}K_{k}
+\frac{1}{3}K_{i} K_{j})\right)
\,{\rm e}^{K(Y,\bar Y)/6}\non~.
\eea
Such a choice implies that
$\cD_{\a}\tilde{\U}\arrowvert
=\cD^{2}\tilde{\U}\arrowvert=0$,
and thus the second line of (\ref{compexpansion})
vanishes. The action reduces to
\be
S_{\rm Kahler} = S_{\rm SG,old}
-3\int\!\!{\rm d}^4x\,e^{-1}\Big\{
-\frac{1}{16}\cD^{2}{\bar \cD}^{2}
\tilde{\U} \arrowvert
+\frac{\rm i}{8}({\bar \J}^{a}{\tilde \s}_{a})^{\a}
\cD_{\a}{\bar \cD}^{2}\tilde{\U} \arrowvert
\Big\}~.
\ee
We are now in a position to write down the component action
for supergravity coupled to a K\"ahler sigma model. 
This result,  which is in agreement 
with previous considerations
(see, e.g.,  \cite{WB}), is
\bea
S_{\rm Kahler} ~=~\int\!\!{\rm 
d}^4x\,e^{-1}&&\!\!\!\Bigg\{\frac{1}{2}\cR
~+~\frac{4}{3}{\Bbb A}\!^{a}{\Bbb A}_{a} ~-~ 
\frac{1}{3}{\bar 
{B}} B
~+~ \frac{1}{4}\ve^{abcd} ({\bar \J}_{a} {\tilde \s}_{b} {\hat \J}_{cd}
- \J_{a} \s_{b} {\hat {\bar \J}}_{cd})\non\\
&&~-~ g_{i{\underline i}}\Big( 
\nabla^{a}Y^{i}\,\nabla\!_{a}{\bar Y}^{\underline i}
~+~\frac{\rm i}{4}(\c^{i} \s^{a}
\! \stackrel{\leftrightarrow}{\hat 
\nabla\!}_{a}
{\bar \c}^{\underline i})
~-~F^{i}{\bar F}^{\underline i}\non\\
&&~-~\frac{1}{2}(\J_{a}\s^{b}{\tilde \s}^{a}\c^{i})
(\nabla\!_{b}{\bar Y}^{\underline i})
~-~\frac{1}{2}({\bar \J}_{a}{\tilde \s}^{b}\s^{a}{\bar \c}^{\underline i})
(\nabla\!_{b}Y^{i}) \\
&&~-~\frac{1}{8}(\J^{a}\s_{b}{\bar \J}_{a})
(\c^{i}\s^{b}{\bar \c}^{\underline i})
~-~\frac{\rm i}{8}\ve^{abcd}(\J_{a}\s_{b}{\bar \J}_{c})
(\c^{i}\s_{d}{\bar \c}^{\underline i}) \Big)\non\\
&&~+~\frac{1}{16}\c^{i}\c^{j}{\bar \c}^{\underline i}{\bar 
\c}^{\underline j}
({\rm R}_{ij{\underline i}{\underline j}}
- \frac{1}{2}g_{i{\underline i}}g_{j{\underline j}})\Bigg\}\non
~,
\eea
where
\bea
{\hat \J}_{ab}{}^\g&=&{\hat \nabla}\!_{a}\J_{b}{}^\g 
- {\hat  \nabla}\!_{b}\J_{a}{}^\g
- {\cT_{ab}}^{c}\J_{c}{}^\g ~,  \non\\
{\hat \nabla}\!_{a}\J_{b}{}^\g 
&=&\nabla\!_{a}\J_{b}{}^\g  ~+~
\frac{1}{4}\left(K_{i}\,\nabla\!_{a}Y^{i} -
K_{\underline i}\,\nabla\!_{a}{\bar Y}^{\underline i}
\right)\J_{b}{}^\g ~,\\
{\hat \nabla}\!_{a}\c^{i}_{\g}&=&\nabla\!_{a}\c^{i}_{\g} ~-~
\frac{1}{4}\left(K_{j}\,\nabla\!_{a}Y^{j} -
K_{\underline j}\,\nabla\!_{a}{\bar Y}^{\underline j}
\right)\c^{i}_{\g}
~+~ {\G^{i}}_{jk} (\nabla\!_{a}Y^{j})\c^{k}_{\g}
~. \non
\eea
In order to  to diagonalize 
in the auxiliary field sector we have
made the redefinition
\be
{A}_{a} ~=~ {\Bbb A}_{a} ~-~ \frac{\rm 
i}{4}\left(K_{i}\,\nabla\!_{a}Y^{i}
- K_{\underline i}\,\nabla\!_{a}{\bar Y}^{\underline i}\right)
~-~ \frac{1}{16}\,g_{i{\underline i}} (\c^{i}\s_{a}{\bar 
\c}^{\underline i})~, 
\ee
so that the auxiliary fields ${\Bbb A}_{a}, {B}$ and $F^{i}$
vanish on the mass shell.

\section{Self-dual electrodynamics in supergravity}
\setcounter{equation}{0}
We are finally prepared to study supersymmetric 
nonlinear electrodynamics.

\subsection{Family of self-dual models}
In \cite{KM}  we constructed a family 
of self-dual models for the Abelian 
vector multiplet  in curved superspace
with actions of the general form
\be
S [W, {\bar W}] =
\frac14\, \intss\, \ERc\, W^2 +
\frac14\, \intss\, \ERac\, {\bar W}^2
+  \frac14\, \intss\, E^{-1} \, W^2\, {\bar W}^2\,
\L(\o, {\bar \o})~,
\label{eq:family action}
\ee
where $\L(\o,\bar \o)$ is a real analytic function of the complex  
variable
\be
\o ~ \equiv ~ \frac{1}{8} \Dsqc\, W^2~.
\label{u}
\ee
Here $ {\bar W}_\ad $ and $W_\a$ are 
covariantly (anti) chiral  superfield 
strengths, 
\be
W_\a = -\frac{1}{4}\, \Dsqac \cD_\a \, V~, \qquad \quad
{\bar W}_\ad = -\frac{1}{4}\, \Dsqc \cDB_\ad \, V ~,
\label{eq:w-bar-w}
\ee
defined in terms of a real unconstrained 
prepotential $V$. The theory (\ref{eq:family action})
is self-dual if  the interaction $\L(\o,\bar \o)$ 
satisfies the following differential equation 
\be
{\rm Im} \,\Big\{ \G
- \bar{\o}\, \G^2
\Big\} = 0~, \qquad \quad
\G  = \frac{\pa (\o \, \L) }{\pa \o}~.
\label{eq:differential}
\ee
The self-dual dynamical systems described
are a curved-superspace generalization 
of the globally supersymmetric systems 
introduced in \cite{KT,KT2}.

To obtain a super-Weyl invariant extension 
of (\ref{eq:family action}), we first note that $V$ 
is inert under  the super-Weyl transformation 
(\ref{superweyl}), and therefore 
the chiral strength $W_\a $ 
transforms as follows
\be
W_\a ~\to ~ {\rm e}^{-3  \s /2} \, W_\a ~.
\ee
Now, implementing 
the substitution (\ref{convertion})
in $S[W, {\bar W}]$, 
with the aid of (\ref{quadr}),  
we then obtain
the super-Weyl invariant action 
\be
S[  W, {\bar W}, \S , {\bar \S}]  = 
S[  W, {\bar W}, \S  {\bar \S}]  ~,
\ee
where 
\bea
S[W,{\bar W}, \U] =
\frac{1}{4}\intss\, \ERc\, W^2 &+&
\frac{1}{4}\intss\, \ERac\, {\bar  W}^2
\non \\
&+&
\frac14\, \intss\, E^{-1} \,
\frac{W^2\,{\bar W}^2}{\U^2}\,
\L\!\left(\frac{\o}{\U^2},
\frac{\bar \o}{\U^2}\right)~.
\label{SED-NSG}
\eea

\subsection{Coupling to K\"ahler 
sigma-models}
In the presence of a  nonlinear 
K\"ahler sigma-model, 
the simplest approach to obtain 
a super-Weyl and K\"ahler
invariant formulation of dynamics 
is to proceed within the framework 
of  new minimal supergravity.
Consider the supergravity-matter system 
described by the action \cite{KM}
\be
S[W,{\bar W}, \f, {\bar \f},{\Bbb L}]
= 3 \intss\, E^{-1}\,
{\Bbb L}\, {\rm ln} {\Bbb L} + \intss\, E^{-1}\,
{\Bbb L}\,K(\f, \bar \f )
+ S[W,{\bar W},{\Bbb L}] ~,
\label{NSG-ED-sigma}
\ee
where $S[W,{\bar W},{\Bbb L}] $ is obtained 
from (\ref{SED-NSG}) by replacing 
$\U \to {\Bbb L}$.
This theory possesses several important symmetries:
(i) super-Weyl invariance; (ii) K\"ahler invariance;
(iii) duality invariance.

To uncover the description of this theory
in the framework of old minimal supergravity, 
let us replace the action (\ref{NSG-ED-sigma})
by the following auxiliary action
\be
S[W, {\bar W}, \f, {\bar \f}, {\Bbb L}, U] =
3 \intss\, E^{-1} \left(U\, {\Bbb L} - \U\right)
+ S[W,{\bar W}, \U] ~,
\label{eq:auxiliary}
\ee
where
\be
\U  = {\rm exp}\!\left(U - \frac13
K(\f, {\bar \f})\right)~.
\label{eq:upsilon}
\ee
Here the additional dynamical variable $U$
is an unconstrained real scalar
superfield.
Varying $U$ brings us back to (\ref{NSG-ED-sigma}).
On the other hand, the equation of motion for 
$\Bbb L$ implies that $U$ takes the form 
(\ref{Usolution}). We thus end up with the action
\bea
S[W, {\bar W}, \f, {\bar \f}, \S, \bar \S] &=&
- 3 \intss\, E^{-1}\,  \tilde{\U}
+ S[W,{\bar W}, \tilde{\U}] ~, 
\label{U(1)}\\
{\tilde \U} &=& \S{\bar \S} \,
{\rm exp}\!\left(-\frac{1}{3} K\!(\f,{\bar \f})\right)~.
\non 
\eea

\subsection{Coupling to the dilaton-axion multiplet}
As demonstrated in \cite{KM},
the supergravity-matter system (\ref{U(1)})
enjoys invariance under electromagnetic 
duality rotations which do not act on the 
supergravity prepotentials and sigma-model 
fields.  The duality group can be shown to be 
${\rm U}(1)$. Building on the ideas developed, 
in particular, in \cite{GR,GZ,KT2}, one can enhance 
the duality group to ${\rm SL}(2,{\Bbb R})$
by coupling the vector multiplet in 
(\ref{U(1)})
to the  dilaton-axion multiplet 
that transforms under duality rotations.
The dilaton-axion complex is described by 
a covariantly chiral scalar superfield, $\F$, 
and takes its values  
in the K\"ahler manifold 
${\rm SL}(2,{\Bbb R})/{\rm U}(1)$. 
This program was explicitly realized in \cite{KM}.
The super-Weyl invariant extension of the
action given in \cite{KM} is 
\bea
\label{eq:d-axcoupledaction}
S &=& - 3 \intss\, E^{-1}\,{\tilde {\bf \U}}
+ \frac{\rm i}{4}\intss\, \ERc\, \F\,W^2 -
\frac{\rm i}{4}\intss\, \ERac\,
{\bar \F}\,{\bar  W}^2
\\
&&-  \frac{1}{16}\, \intss\, E^{-1} \,
(\F-{\bar \F})^2\, \frac{W^2\,{\bar W}^2}{{\tilde {\bf \U}}^2 }  \,
\L \Big( 
\frac{\rm i}{2}(\F-{\bar \F})\,\frac{\o}{{\tilde {\bf \U}}^2}\;,\;
\frac{\rm i}{2}(\F-{\bar \F})\,\frac{\bar \o}{{\tilde {\bf \U}}^2} 
\Big)~,\non
\eea
where
\be
{\tilde {\bf \U}}=\S{\bar \S}
\exp\!\Big(-{1 \over 3}\cK\!(\F, \bar \F )
-{1 \over 3} K\!(\f, \bar \f) \Big)~.
\ee
Here $\cK(\F,{\bar \F})$
denotes the  K\"ahler potential
of the  manifold 
${\rm SL}(2,{\Bbb R})/{\rm U}(1)$. 
It has the form 
\be
\label{eq:kahlerpotential}
\cK(\F,{\bar \F})=-{\rm ln}\,{ {\rm i} \over 2}
(\F-{\bar \F})~.
\ee
The action (\ref{U(1)}) follows from 
(\ref{eq:d-axcoupledaction}) by setting 
$\F = - {\rm i}$.

Now, it is our aim to 
analyze the component structure 
of the theory with action (\ref{eq:d-axcoupledaction}).

\subsection{Component reduction}
We proceed by introducing  
the component fields of the vector multiplet
\bea
W_{\a}\arrowvert = \j_{\a}~,\qquad
-\frac{1}{2}\cD^{\a}W_{\a}\arrowvert=D~,
\qquad
\cD_{(\a}W_{\b)}\arrowvert 
=2 {\rm i} {\hat F}_{\a\b}
&=& {\rm i} (\s^{ab})_{\a\b}{\hat F}_{ab}~,
\eea
where
\bea
\label{eq:F_ab defn}
{\hat F}_{ab} &=& F_{ab} -
\frac{1}{2}(\J_{a}\s_{b}{\bar \j} + 
\j\s_{b}{\bar \J}_{a}) +
\frac{1}{2}(\J_{b}\s_{a}{\bar \j} + 
\j\s_{a}{\bar \J}_{b})~,
\non\\
F_{ab} &=& \nabla\!_{a}V_{b} 
- \nabla\!_{b}V_{a} 
- {\cT_{ab}}^{c}V_{c}~,
\eea
with $V_a= e_a{}^m (x) \,V_m (x)$  
the  gauge  one-form.

Similarly to our definition (\ref{scalarcompdef}) of 
the component fields $\left\{Y^{i},\c^{i}_{\a},F^{i}\right\}$ 
of  the scalar superfield $\f^{i}$, we 
introduce the component 
fields $\left\{\cY,\eta_{\a},\cF\right\}$ of the dilaton-axion 
multiplet $\F$. The dilaton\footnote{We have used
the same  Greek letter $\vf$ to denote 
the  chiral prepotential 
and the dilaton.
Only the latter
occurs in the remainder of this paper.} 
$\vf$ and axion $a$ fields are 
related to the superfield $\F$ by
\be
\F\arrowvert=\cY=a-{\rm i}\,{\rm e}^{-\vf}~.
\ee

Applying the reduction rule 
(\ref{reductionrule}) to the action 
(\ref{eq:d-axcoupledaction}) we obtain
\bea
\label{eq:d-axcoupledaction2}
S=S_{V}\!&-&\!3\int\!\!{\rm d}^4x\,e^{-1}\Big\{\Big(
-\frac{1}{4}\cD^{2}R\arrowvert-
\frac{\rm i}{2} ({\bar \J}^{a}{\bar 
\s}_{a})^{\a}\cD_{\a}R\arrowvert
+ (B + {\bar \J}^{a}{\tilde \s}_{ab}{\bar \J}^{b}) R\arrowvert
\Big)
\O\arrowvert\non\\
&&\qquad-\frac{1}{2}\cD^\a R\arrowvert
\cD_{\a}\O\arrowvert
-\frac{1}{4}R\arrowvert
\cD^2\O\arrowvert
-\frac{\rm i}{2}({\bar \J}^{a}{\tilde \s}_{a})^{\a}
\cD_{\a}\O\arrowvert\\
&&\qquad
+\frac{1}{16} \cD^2{\bar \cD}^2\O\arrowvert
+\frac{\rm i}{8}({\bar \J}^{a}{\tilde \s}_{a})^{\a}
\cD_{\a}{\bar \cD}^2\O\arrowvert
-\frac{1}{4}(B+{\bar \J}^a{\tilde \s}_{ab}{\bar \J}^b)
{\bar \cD}^2\O\arrowvert
\Big\}~,\non
\eea
where
\be
\O={\tilde {\bf \U}}+
\frac{1}{48}(\F-{\bar \F})^2\frac{W^2{\bar W}^2}{\tilde{\bf \U}^2}\,
\L\Big(\frac{\rm i}{2}(\F-{\bar \F})\,\frac{\o}{{\tilde {\bf \U}}^2} \;, \;
\frac{\rm i}{2}(\F-{\bar \F})\,\frac{\bar \o}{{\tilde {\bf \U}}^2}\Big)~,
\ee
and we have separated out the following part of the action:
\be
S_{V}=\frac{\rm i}{4}\intss\, \ERc\,\F\,W^2 -
\frac{\rm i}{4}\intss\, \ERac\,
{\bar \F}\,{\bar  W}^2~.
\ee
Since $S_{V}$ does not couple to $\S$ and $\bar \S$, 
its component form  is independent 
of the super-Weyl gauge choice. 
It is therefore straightforward to evaluate 
the component structure of this part of the action
\bea
S_{V} &=& \int\!\!{\rm d}^4x\,e^{-1} \Bigg\{
- \frac{1}{4}\,{\rm e}^{-\vf}\,F^{ab}F_{ab}
+\frac{1}{4}\,a\,F^{ab}{\tilde F}_{ab}
-\frac{1}{2}(\j\s^{b}{\bar \j})\nabla\!_{b}a
-\frac{\rm i}{2}\,{\rm e}^{-\vf}\,(\j\s^{a}\nabla\!_{a}{\bar \j})\non\\
&&+\frac{1}{2}\,{\rm e}^{-\vf}\,F^{ab}
(\J_{a}\s_{b}{\bar \j} + \j\s_{b}{\bar \J}_{a})
+\frac{\rm i}{2}\,{\rm e}^{-\vf}\,{\tilde F}^{ab}
(\J_{a}\s_{b}{\bar \j} - \j\s_{b}{\bar \J}_{a})\non\\
&&+\frac{1}{4}\,F^{ab}(\eta\s_{ab}\j
+{\bar \eta}{\tilde \s}_{ab}{\bar \j})
+\frac{1}{4}\left({\rm e}^{-\vf}
(\eta^{ac}\eta^{bd}-\eta^{ad}\eta^{bc})
+a\,\e^{abcd}\right)
(\J_{a}\s_{b}{\bar \j})(\j\s_{c}{\bar \J_{d}})\non\\
&&+\frac{1}{16}\,{\rm e}^{-\vf}
\Big((3{\bar \J}^{a}{\bar \J}_{a}
-2{\bar \J}^{a}{\tilde \s}_{ab}{\bar \J}^{b})\j^{2}
+(3\J^{a}\J_{a}
-2\J^{a}\s_{ab}\J^{b}){\bar \j}^{2}
\Big)\non\\
&&-\frac{1}{8}(\J_{a}\s_{b}{\bar \j})
(\eta\s^{ab}\j)
-\frac{1}{8}({\bar \J}_{a}{\tilde \s}_{b}\j)
({\bar \eta}{\tilde \s}^{ab}{\bar \j})
-\frac{1}{32}\j^2(\eta\s^{a}{\bar \J}_{a})
+\frac{1}{32}{\bar \j}^2(\J_{a}\s^{a}{\bar \eta})
\non\\
&&+\frac{1}{16}\,{\rm e}^{\vf}\,(\eta^2\j^2
+{\bar \eta}^{2}{\bar \j}^{2})
+\frac{1}{2}(\j\s^{a}{\bar \j}){\cT_{ab}}^{b}
-\frac{\rm i}{4}(\eta\j-{\bar \eta}{\bar \j})D\\
&&+\frac{1}{2}\,{\rm e}^{-\vf}\,D^{2}
-{\rm e}^{-\vf}
(\j\s^{a}{\bar \j})A_{a}
+\frac{\rm i}{4}(\cF\j^2-{\bar \cF}{\bar \j}^2)
\Bigg\}~,\non
\eea
where we have used the explicit form of the K\"ahler
potential (\ref{eq:kahlerpotential}). This result is 
in agreement with \cite{BGG}.

Looking at the first line of (\ref{eq:d-axcoupledaction2}) 
we notice that if a super-Weyl gauge choice is made such 
that $\O\arrowvert=1$ then this will reduce to the supergravity 
action (\ref{compsugra}), and not 
require a Weyl rescaling. 
To achieve this, we make the choice 
\be
\S\arrowvert = 
{\rm exp}\!\left(\frac{1}{6}K\!(Y,\bar Y) + 
\frac{1}{6}\cK\!(\cY,\bar \cY)
+\frac{1}{24}
{\rm e}^{-2\vf}\j^2{\bar \j}^2 
\L\!\!\left({\rm e}^{-\vf}\o\arrowvert\;,\; 
{\rm e}^{-\vf}{\bar \o}\arrowvert\right)
\right)~.
\ee
A number of options are available for the gauge choice 
for the other components of $\S$. If the following gauge 
choices are made
\bea
&&\quad
\cD_\a\S\arrowvert=\frac{1}{3}
(\c^{i}_{\a}K_{i}-\frac{\rm i}{2}\,{\rm e}^{\vf}
\eta_\a)\,\S\arrowvert~,\non\\
-\frac{1}{4}\cD^2 \S\arrowvert &=& 
\frac{1}{3}\Big(F^{i} K_{i}
- \frac{1}{4}\,\c^{i}\c^{j} (K_{ij} - {\G^{k}}_{ij}K_{k}
+\frac{1}{3}K_{i} K_{j}) \\
&&\quad\quad\qquad
-\frac{\rm i}{2}\,{\rm e}^{\vf}\cF
-\frac{1}{24}\,{\rm e}^{2\vf}\,\eta^2 
+\frac{\rm i}{12}\,{\rm e}^{\vf}\,\eta\c^{i}K_{i}
\Big)\,\S\arrowvert~,\non
\eea
then $\cD_{\a}{\tilde {\bf \U}}\arrowvert=
\cD^{2}{\tilde {\bf \U}}\arrowvert=0$, and 
the action (\ref{eq:d-axcoupledaction2}) 
simplifies greatly.

The complete component action turns out to be 
extremely complicated as far as the fermionic sector 
is concerned.  The fermionic sector will 
be studied in the flat-space case in sections 5 and 6. 
Here we only  focus on the 
bosonic sector.
\bea
\label{eq:bosonic action}
S_{\rm bosonic} &=& \int\!\!{\rm d}^4x\,e^{-1} \Bigg\{~
\frac{1}{2} \cR 
- g_{i{\underline i}}\,
\nabla^{a} 
Y^{i}\,\nabla\!_{a}{\bar Y}^{\underline i}
-\frac{1}{4}\Big( {\rm e}^{2\vf}\,
(\nabla a)^2
+  (\nabla \vf )^2  \Big)
\non \\
&&\quad\qquad\qquad
-\frac{1}{4}\,{\rm e}^{-\vf}F^{ab}F_{ab} 
+\frac{1}{4}\,a\,F^{ab}{\tilde F}_{ab}
+{\rm e}^{-2\vf}\,w{\bar w}\,\L\!\left(
{\rm e}^{-\vf}w\,,\,{\rm e}^{-\vf}{\bar w}\right)\\
&&\quad\qquad\qquad
+ \frac{4}{3}{\Bbb A}\!^{a}{\Bbb A}_{a} 
- \frac{1}{3}B{\bar B}
+ \hf {\rm e}^{-\vf}D^{2}
+ g_{i{\underline i}} \,
F^{i}{\bar F}^{\underline i}
+\frac{1}{4}\,{\rm e}^{2\vf}\,\cF{\bar \cF}~
\Bigg\}\non~,
\eea
where
\bea
w&=&F^{\a\b}F_{\a\b}-\hf D^2~,\quad\qquad
{\bar w}={\bar F}^{\ad\bd}{\bar F}_{\ad\bd}-\hf D^2~,\non\\
A_{a} &=& {\Bbb A}_{a} 
-\frac{\rm i}{4}\left( K_i \,
\nabla\!_{a}Y^{i}
- K_{\underline i}\,
\nabla\!_{a}{\bar Y}^{\underline i} \right)
-\frac{1}{4}\,{\rm e}^{\vf} \,\nabla\!_a a
~,
\eea
and $\cR$ and $F_{ab}$ are as defined respectively 
in (\ref{eq:covariant derivative algebra}) and 
(\ref{eq:F_ab defn}), but with torsion set to zero.

As a special representative in  
the family of self-dual actions 
(\ref{eq:family action})--(\ref{eq:differential}), 
we would like to consider 
the supersymmetric Born-Infeld action. In this case 
the function $\L(\o,{\bar\o})$ takes the form
\bea
&&\L(\o,{\bar\o}) =
\frac{\k^2}
{ 1 + \hf\, A \, + \sqrt{1 + A +\frac{1}{4} \,B^2} }~,\\
&&\quad
A=\k^2(\o  + \bar \o)~, \qquad 
B=\k^2(\o - \bar \o)~.\non
\eea
After eliminating the auxiliary fields, the bosonic action  
(\ref{eq:bosonic action}) becomes
\bea
\label{BI-curved}
S &=& \int\!\!{\rm d}^4x\,e^{-1} \Bigg\{\,
\frac{1}{2}\cR
- g_{i{\underline i}}\,
\nabla^{a} 
Y^{i}\,\nabla\!_{a}{\bar Y}^{\underline i}
-\frac{1}{4}\Big( {\rm e}^{2\vf}\,
(\nabla a )^2 + (\nabla \vf )^2 \Big)
\non\\
&&\quad\qquad\qquad
+\frac{1}{\k^2}\left(1-\sqrt{-{\rm det}(\eta_{ab}
+\k\,{\rm e}^{-\vf/2}F_{ab})}\right)
+\frac{1}{4}\,a\,F^{ab}{\tilde F}_{ab}
\Bigg\}~.
\eea

\section{Photino dynamics in flat space}
\setcounter{equation}{0}
Our discussion of the fermionic dynamics 
in $\cN=1$ supersymmetric nonlinear electrodynamics 
will be restricted to the case of flat global superspace.
Here the action takes the form 
\be
S [W, {\bar W}] =
\frac14 \int {\rm d}^6 z \, W^2 +
\frac14 \int {\rm d}^6 {\bar z} \, {\bar W}^2
+  \frac14 \intss\,   W^2\, {\bar W}^2\,
\L(\o, {\bar \o})~,
\label{family}
\ee
where 
\be
\o ~ \equiv ~ \frac{1}{8} D^2 W^2~.
\ee
If $\L(\o, {\bar \o})$ is a solution 
of  the equation (\ref{eq:differential}), 
then the above action 
obeys the self-duality equation \cite{KT,KT2,KM}
\be 
{\rm Im} \int {\rm d}^6 z \, \Big\{ 
W^2 + M^2 \Big\} =0 ~, 
\qquad 
{ {\rm i} \over 2} \,M_\a  = 
{\d  \over \d W^\a} \,S [W, {\bar W}]  ~.
\label{sde-flat}
\ee

The action (\ref{family}) can be seen to be  invariant under
a discrete chiral transformation 
\be
W_\a (x, \q) ~ \longrightarrow ~
W_\a (x, -\q)~,
\ee
which leaves the fermionic fields invariant, 
\be 
\j_\a(x) = W_\a | ~ \longrightarrow ~
\j_\a(x)~, 
\ee
whilst changing the bosonic fields as follows:
\be
F_{\a \b} (x) = {1 \over 2{\rm i} } D_{(\a} W_{\b)} |
~ \longrightarrow ~ - F_{\a \b} (x) ~, 
\qquad 
D(x) = -{1 \over 2} D^\a W_\a | 
~ \longrightarrow ~ - D(x)~.
\ee
This symmetry implies that the component action 
contains only even powers of the bosonic fields. 
It is therefore consistent, when discussing 
the component structure,  to restrict our attention 
to the purely fermionic sector specified by 
\be 
D_\a W_\b | =0~. 
\label{ferm-reduc}
\ee
Let $S[\j, {\bar \j}]$ be the fermionic action 
that follows from (\ref{family}) upon switching off 
all the bosonic fields. It turns out that 
 $S[\j, {\bar \j}]$ obeys a functional equation 
which is induced by the self-duality  (\ref{sde-flat}).
 
The self-duality equation (\ref{sde-flat})
must hold for an arbitrary 
chiral spinor $W_\a(z)$ and its conjugate
${\bar W}_\ad (z)$. This means that 
the spinors $W_\a$ and 
${\bar W}_\ad$ are chosen 
in (\ref{sde-flat}) to satisfy 
only the chirality constraints 
${\bar D}_\ad W_\a =0$ and 
$D_\a {\bar W}_\ad =0$, but not the Bianchi 
identity  
\be 
D^\a W_\a = {\bar D}_\ad {\bar W}^\ad ~. 
\ee
Thus $W_\a$ now contains two independent 
fermionic components
\be
\j_\a(x) = W_\a | ~, \qquad 
\r_\a(x) = -{1 \over 4} D^2 W_\a|~.
\ee
Let $\hat{S} \equiv S[\j, {\bar \j}, \r , {\bar \r}]$ be the 
component   action 
that follows from (\ref{family}) upon 
relaxing  the Bianchi identity and restricting to 
the fermionic sector  (\ref{ferm-reduc}). 
Then, the  self-duality equation (\ref{sde-flat})
reduces to 
\be 
{\rm Im} \int {\rm d}^4 x \, \Big\{ 
\j^\a \r_\a  + 4 \, 
\frac{\d \hat{S}}{\d \j^\a} \,
\frac{\d \hat{S}}{\d \r_\a} \Big\} =0~.
\ee
The genuine fermionic action, $S[\j, {\bar \j}]$, 
is obtained from the self-dual action $\hat{S}$ 
by imposing the  ``fermionic Bianchi identities''
$ \r_\a 
= - {\rm i} \,(\s^b \pa_b {\bar \j})_\a$
and ${\bar \r}_\ad =  {\rm i} \,( \pa_b \j \s^b)_\ad\,$,
\be 
S[\j, {\bar \j}] = S[\j, {\bar \j}, \r, 
{\bar \r}]
\Big|_{ \r
= - {\rm i} \,(\s^b \pa_b {\bar \j})}~.
\ee

A short calculation leads to the fermionic action  
\bea
S[\j, {\bar \j}] &=&  \int {\rm d}^4 x\, \Big\{ 
- \hf \la  u+ {\bar u}  \ra 
+ \Big( \la  u \ra \la {\bar u}  \ra 
-{1\over 4}  (\pa^a \j^2) (\pa_a {\bar \j}^2) \Big)\,
\L(0,0) \non \\
&&+ \la  u \ra \Big( \la  u \ra \la {\bar u}  \ra 
+\hf ({\bar \j}^2 \Box \j^2)  \Big) \,\L_\o(0,0)
+  \la {\bar u}  \ra \Big( \la  u \ra \la {\bar u}  \ra 
+\hf (\j^2 \Box {\bar \j}^2)  \Big) \,\L_{\bar \o}(0,0)
\non \\
&&+ \Big( \la  u \ra^2 \la {\bar u}  \ra^2
- \hf  (\pa^a \j^2) (\pa_a {\bar \j}^2) 
\la  u \ra \la {\bar u}  \ra 
+{1 \over 16} \j^2{\bar \j}^2 
(\Box \j^2) ( \Box {\bar \j}^2) \Big) 
\L_{\o{\bar \o}}(0,0) \non \\
&&+ {3\over 8} ({\bar \j}^2 \Box \j^2)
 \la  u \ra^2 \,\L_{\o \o}(0,0)
+{3\over 8} (\j^2 \Box {\bar \j}^2) 
\la {\bar u}  \ra^2 \, 
\L_{{\bar \o}{\bar \o}}(0,0)\Big\}~. 
\label{fac}
\eea
Here we have introduced 
the following $4 \times 4$ matrices: 
\be 
 u_a{}^b = {\rm i} \,\j \s^b \pa_a{\bar \j}~, 
\qquad 
{\bar u}_a{}^b = - {\rm i}\, (\pa_a \j) \s^b {\bar \j}~, 
\ee
as well as made use of  
the useful compact 
notation 
\be
\la F \ra \equiv {\rm tr}\, F = F_a{}^a~,
\label{not}
\ee 
for an arbitrary  $4 \times 4$ matrix $F=(F_a{}^b)$.

The fermionic action  obtained involves 
several constant parameters associated with the function 
$\L(\o,{\bar \o}) $ that enters the original supersymmetric 
action. However, not all of these parameters are 
independent  since 
$\L(\o,{\bar \o}) $ must be a solution 
to the self-duality equation (\ref{eq:differential}). 
This restriction proves to imply 
\be 
\L_\o(0,0)= \L_{\bar \o}(0,0) =- \L^2(0,0)~, 
\qquad 
\L_{\o \o}(0,0) = \L_{{\bar \o}{\bar \o}}(0,0)
= 2\L^3(0,0)~.
\label{sel-d-con-1}
\ee
The self-duality equation imposes no restrictions on 
$ \L(0,0)$ and $\L_{\o{\bar \o}}(0,0) $.
${}$For later convenience, we represent 
\be 
\L(0,0) = \frac{\k^2}{2}~,\quad \qquad~
\L_{\o{\bar \o}}(0,0)=\frac{\k^{6}}{8}(\m+3)~. 
\label{sel-d-con-2}
\ee

\section{Relation to the Akulov-Volkov action}
\setcounter{equation}{0}
Looking at the fermionic action (\ref{fac}), 
it is hardly possible to imagine that it is 
related somehow to the Akulov-Volkov action 
(\ref{AV}), which describes Goldstino 
dynamics \cite{VA,AV}
and which can be represented in the form
\bea
S_{\rm AV}[\l, {\bar \l}] 
&=&  -{1 \over2} \int {\rm d}^4 x \, \Big \{
\la  v+ {\bar v}  \ra 
+\frac{\k^2}{2}  \Big(
\la v  \ra \la {\bar v} \ra
- \la v  {\bar v} \ra \Big)
\non \\
&&+\frac{\k^4}{16}  \Big( 
\la  v^2 {\bar v} \ra
-\la v  \ra \la  v {\bar v} \ra
-\hf \la  v^2 \ra \la  {\bar v}  \ra
+ \hf \la v  \ra^2 \la {\bar v } \ra  
~+~ {\rm c.c.} \Big)  \Big\}~,
\label{AV3}
\eea
see Appendix A for more details.
Here 
\bea
v_a{}^b = 
{\rm i} \,\l \s^b \pa_a{\bar \l}  ~,
\qquad 
{\bar v}_a{}^b=
- {\rm i}\, (\pa_a \l) \s^b {\bar \l}~. 
\eea
Nevertheless, the two fermionic theories 
turn out to be closely related in the following sense. 
There exists a nonlinear field redefinition, 
$(\j_\a , {\bar \j}^\ad ) \to   (\l_\a , {\bar \l}^\ad )$, 
that eliminates all the higher derivative terms 
in (\ref{fac}) and brings this action to 
a one-parameter deformation of the AV action. 
The two theories coincide, modulo 
such a field redefinition,  under the choice 
\be
\L_{\o{\bar \o}}(0,0) = \frac{3}{8} \k^{6} = 3 \L^{3}(0,0) 
\quad \Longleftrightarrow \quad \m=0~,
\label{cond-for-AV}
\ee
which occurs, in particular, 
in the case of the supersymmetric Born-Infeld
action \cite{CF,BG,RT}
\bea
S_{\rm SBI} &=&
\frac14 \int {\rm d}^6 z \, W^2 +
\frac14 \int {\rm d}^6 {\bar z} \, {\bar W}^2
+\frac{\k^2}{4} \intss\,\frac{W^2\, {\bar W}^2}
{ 1 + \hf\, A \, + \sqrt{1 + A +\frac{1}{4} \,B^2} }~,
\label{born-infeld2} \\
&& A=\k^2\,(\o  + \bar \o)~, \qquad 
B = \k^2\,(\o - \bar \o)~, 
\qquad 
\o = \frac{1}{8} D^2 W^2~.
\non
\eea
This section is devoted to the proof of 
the above statement.

We begin looking for a field redefinition 
by first noting that 
the leading order terms must match, 
$\psi_{\a} = \l_{\a} + O(\k^{2})$. 
Next, to third-order in fields the general form 
of the redefinition can be written as
\be
\psi_\a = \l_{\a}\Big\{1+ 
\frac{\k^{2}}{2} \a_{1} \la v \ra + 
\frac{\k^{2}}{2} \a_{2} \la \bar{v} \ra \Big\}+ 
\frac{{\rm i}\k^{2}}{2} \a_{3} (\s^{a}\bar{\l})_{\a} (\pa_{a}\l^{2})
 + O(\k^{4})~,
\label{cub-red}
\ee
where the constant coefficients $\a_{1}, \a_{2}, \a_{3}$  
can be chosen to be real. 
Substituting (\ref{cub-red}) into (\ref{fac}) gives
\bea
S[\j, {\bar \j}] 
&=&  -{1 \over2} \int {\rm d}^4 x \, \Big \{
\la  v+ {\bar v}  \ra 
+\k^{2} \a_{1} \left(\la v \ra^{2} + \la {\bar v} \ra^{2}\right)
+2\k^{2} (\a_{2}+\a_{3}-\frac{1}{2}) \la v \ra \la \bar{v} \ra\non\\
&& - 2\k^{2} \a_{3} \la v {\bar v} \ra
-\k^{2}(\a_3-\frac{1}{4}) (\pa^{a}\l^2) (\pa_{a} \bar{\l}^2)
\Big\} + O(\k^{4})~.
\eea
The requirement that the transformed action 
match with the AV action 
(\ref{AV3}), uniquely fixes the coefficients 
$\a_{1} = 0, \a_{2} = \frac{1}{2}, \a_{3} = \frac{1}{4}$. 
A similar calculation at  fifth-order also allows us 
to match the AV action to this order. 
However, this calculation proves to be  
extremely tedious, as there 
exist many more admissible structures 
that can contribute to  the field redefinition 
under consideration. Unlike the 
third-order case, not all coefficients 
are uniquely fixed -- 
we are left with three free parameters, 
$\b_{1}, \b_{2}, \b_{3}$. 
At the highest-order, this is again the case, 
and we gain another free parameter, $\g$. 
However, at this order, 
even with this freedom in the redefinition, 
it is impossible to match 
the AV action unless a restriction is placed 
on the type of model we are investigating, 
i.e. we must choose a particular value for 
$\L_{\o{\bar \o}}(0,0)$.

With the following field redefinition
\bea	
\psi_\a &=& \l_{\a}\Big\{1+\frac{\k^{2}}{4} \la \bar{v} \ra
\label{field-redef}
\\
&& \phantom{ \l_{\a}\Big\{1 }
+ \frac{\k^{4}}{4} \Big(\b_{1} \la v \ra \la \bar{v} \ra
+ \b_{2} \la \bar{v} \ra^{2}
+ (2\b_{3}-\frac{1}{4}) \la v\bar{v} \ra
-\frac{1}{4} \la \bar{v}^{2} \ra
+ \b_3 (\pa^{a}\l^2) (\pa_{a} \bar{\l}^2)
+ \frac{1}{16} (\bar{\l}^{2} \Box \l^{2}) \Big)\non\\
&& \phantom{ \l_{\a}\Big\{1 }
+\frac{\k^{6}}{64} \Big( (3 \m+1+4(\b_{1}-2\b_{3}-2\g))
\la v \ra \la \bar{v} \ra^{2}
-2(\m-2\b_{1}-2\b_{2})
\la \bar{v} \ra (\pa^{a}\l^2) (\pa_{a} \bar{\l}^2)\non\\
&&  \phantom{ \l_{\a}\Big\{1 }
\qquad \quad ~\,
+ (\m - \frac{1}{4}-2\b_{1}+4\b_{3})
\la v \ra (\bar{\l}^{2} \Box \l^{2})
+ 8(\b_{1} + \b_{2} - \b_{3}) \la v\bar{v}^{2} \ra
\Big)\Big\} \non\\
&&+ \frac{\rm{i}}{8}\k^2 (\s^{a}\bar{\l})_{\a} (\pa_{a}\l^{2})
 \Big\{
1 + \frac{\k^{2}}{2} (1 - 4(\b_{1}+\b_{2}+\b_{3})) \la v \ra
+ 2\k^{2} \b_{3} \la \bar{v} \ra
+ \k^{4} \g \la v \ra \la \bar{v} \ra
\Big\}~,\non
\eea
the transformed action is
\bea
S[\j, {\bar \j}] = S_{\rm AV}[\l, {\bar \l}] + 
\frac{\k^{6}}{32}\,\m \int {\rm d}^4 x
\, \la v^2 \bar{v}^2 \ra ~.
\label{fac2}
\eea
We see that for the action (\ref{fac}),
in conjunction with (\ref{sel-d-con-1})
and (\ref{sel-d-con-2}), 
is equivalent to 
the AV action (\ref{AV3}) 
if eq.  (\ref{cond-for-AV}) holds.
In particular, the supersymmetric Born-Infeld action 
(\ref{born-infeld2})  has this property.

Our field redefinition (\ref{field-redef}) 
involves four free parameters, 
$\b_{1}, \b_{2}, \b_{3}$ and  $\g$,
which do not show up in the transformed 
action (\ref{fac2}). 
This means that these parameters 
correspond  to some  
symmetries of the original theory (\ref{fac}).
Indeed, if we modify the field redefinition 
(\ref{field-redef}) by  varying any of the parameters, 
the action  is not affected.

\section{Discussion}
\setcounter{equation}{0}
At first glance, the existence of the field redefinition
(\ref{field-redef})  that turns the action (\ref{fac})
 into (\ref{fac2}), looks absolutely fantastic
and unpredictable. 
However, it has a solid
theoretical justification in one special case of
self-dual supersymmetric electrodynamics 
(\ref{family}) -- the $\cN =1$
supersymmetric Born-Infeld action 
(\ref{born-infeld2}). This action
is known to describe the Goldstone-Maxwell multiplet for
spontaneous partial supersymmetry breaking
$\cN=2 \to \cN=1$ \cite{BG,RT}.
As a consequence, its purely fermionic sector
\bea
S_{\rm BG}[\j, {\bar \j}] &=&  \int {\rm d}^4 x\, \Big\{ 
- \hf \la  u+ {\bar u}  \ra 
+ \frac{\k^2}{2}
 \Big( \la  u \ra \la {\bar u}  \ra 
-{1\over 4}  (\pa^a \j^2) (\pa_a {\bar \j}^2) \Big)\,
\non \\
&&-\frac{\k^4}{4} \Big( \la  u \ra \Big[\la  u \ra \la {\bar u}  \ra +\hf ({\bar \j}^2 \Box \j^2) \Big]
+ \la {\bar u}  \ra \Big[ \la  u \ra \la {\bar u}  \ra 
+\hf (\j^2 \Box {\bar \j}^2)  \Big]\Big) 
\non \\
&&+ \frac{3\k^6}{32} \Big( 4\la  u \ra^2 \la {\bar u}  \ra^2
- 2  (\pa^a \j^2) (\pa_a {\bar \j}^2) 
\la  u \ra \la {\bar u}  \ra 
+{1 \over 4 } \j^2{\bar \j}^2 
(\Box \j^2) ( \Box {\bar \j}^2) 
\non \\
&&
\phantom{+ 6 \k^2 \Big( 4\la  u \ra^2 \la {\bar u}  \ra^2}
+ ({\bar \j}^2 \Box \j^2)
 \la  u \ra^2 
+ (\j^2 \Box {\bar \j}^2) 
\la {\bar u}  \ra^2 \, 
\Big)\Big\}~, 
\label{fac3}
\eea
which follows from (\ref{fac}),
turns out to describe
spontaneous breakdown of $\cN=1$
 supersymmetry \cite{HK}.
This Goldstino action clearly does not coincide
with the standard Goldstino action (\ref{AV}) or,
equivalently, with (\ref{AV3}). 
Universality of the Goldstino dynamics, 
on the other hand,  implies that the
two Goldstino actions, (\ref{AV}) and (\ref{fac3}),
should be related to each other.
It was therefore conjectured in \cite{HK} that 
the actions (\ref{AV}) and (\ref{fac3})
 are related by a nontrivial
field redefinition. Moreover, guided by considerations
of nonlinearly realized supersymmetry, 
the authors of \cite{HK}
proposed a nice scheme for constructing 
such a field redefinition and 
also  confirmed it to order $\k^2$
(see the first line in (\ref{field-redef})).
Pushing their scheme to higher orders 
seems to give the redefinition 
(\ref{field-redef}) with all the parameters
fixed as follows: 
$\b_{1} = \frac{1}{16},$ $\b_{2} = 0$, 
$\b_{3} = \frac{1}{32}$ and  $\g = 0$. 
We have checked the correspondence to 
order $\k^4$.

The field redefinition (\ref{field-redef})  
corresponds to the purely fermionic sector of 
the globally supersymmetric theory (\ref{family}). 
In the case when both bosonic and 
fermionic fields are present, 
as well as in the presence of supergravity
-- the case we analyzed in section 4,
there should 
exist an extension of (\ref{field-redef}) 
that, at least, eliminates all higher derivative 
terms from the component action.
But here our brute-force approach 
becomes extremely  cumbersome and 
tedious to follow (even the fermionic
case was quite a pain). 
We believe that there should be  a more 
efficient approach to construct such 
field redefinitions. Unfortunately, 
it is beyond our grasp at the moment.
It is worth pointing out that the issue 
of constructing nonlinear field redefinitions
that eliminate higher derivatives, 
is quite typical in supersymmetric 
field theories. It naturally occurs when 
studying low-energy effective actions
in extended super Yang-Mills theories
\cite{GKPR,KM2}. 

In conclusion, we would like to make
a final comment regarding the supersymmetric 
Born-Infeld action (\ref{born-infeld2}). 
In the purely bosonic sector, 
this theory reduces, upon elimination
of the auxiliary field, to the Born-Infeld action
\be
\label{BI}
S_{\rm BI}= \frac{1}{\k^2} \int {\rm d}^4x \,
\left(1-\sqrt{-{\rm det}(\eta_{ab}
+\k\,F_{ab})}\right)~,
\ee
compare with (\ref{BI-curved}).
In the purely fermionic sector, 
it reduces, upon implementing 
the field redefinition (\ref{field-redef})  
with $\m=0$, to the Akulov-Volkov 
action (\ref{AV}).
In the general case, it should describe,
upon implementing a nonlinear field redefinition,
the space-time filling D3-brane in a special 
gauge for kappa-symmetry, 
see \cite{APS,Tseytlin} and references therein. 
Such a gauge differs from the one chosen in \cite{APS}.

The supersymmetric Born-Infeld action 
(\ref{born-infeld2}) is just a special representative 
in the family of self-dual models (\ref{family}),
with  $\L(\o, {\bar \o})$  a solution of  
the differential  equation (\ref{eq:differential}).
But it is only the action  (\ref{born-infeld2})
which describes the partial supersymmetry 
breaking $\cN=2 \to \cN=1$.
At the component level, however, 
the purely fermionic action has been shown to be 
equivalent to the Goldstino action (\ref{AV}) 
under the mild condition  (\ref{cond-for-AV}), 
which holds for infinitely many members
of the family, including the supersymmetric 
Born-Infeld action (\ref{born-infeld2}).
In this sense, all such models contain 
information about spontaneous supersymmetry 
breaking. 
\vskip.5cm

\noindent 
{\bf Note added:}\\
We are grateful to Simon Tyler for pointing out typos in eqs. (5.12) 
and (7.1).

\noindent
{\bf Acknowledgements:}\\ 
Discussions with Ian McArthur are 
gratefully acknowledged.
The work of SK  is supported in part by 
the Australian Research Council. 
The work of SMc is supported by 
the Hackett Postgraduate Scholarship 
and a UWA Graduates  Association 
Postgraduate Research Travel Award.

\begin{appendix}

\section{The Akulov-Volkov action}
\setcounter{equation}{0}
The Akulov-Volkov (AV) action for the Goldstino 
\cite{VA} is\footnote{Note that the normalization factor 
used here differs from that of $1/2\k^2$ usually found in the 
literature. This is in order to match up with the 
coupling constant in the bosonic sector of the supersymmetric 
Born-Infeld action (\ref{BI}).}
\bea
S_{\rm AV}[\l, {\bar \l}] =  \frac{2}{\k^2} \int {\rm d}^4 x
\, \Big\{ 1 - \det \Xi \Big\} ~, 
\label{AV}
\eea
where 
\bea
\Xi_a{}^b = \d_a{}^b 
+\frac{\k^2}{4}  \Big({\rm i} \,\l \s^b \pa_a{\bar \l}  
- {\rm i}\, (\pa_a \l) \s^b {\bar \l} \Big) 
\equiv \d_a{}^b + \frac{\k^2}{4}(v +{\bar v})_a{}^b~.
\eea
With the notation (\ref{not}),
the Goldstino action can explicitly be rewritten 
as a polynomial in $v$ and $\bar v$:
\bea 
S_{\rm AV}[\l, {\bar \l}] 
&=&  -{1 \over2} \int {\rm d}^4 x \, \Big \{
\la  v+ {\bar v}  \ra 
+ \frac{\k^2}{8} \Big(
\la v  +{\bar v} \ra^2 
- \la   (v  +{\bar v})^2 \ra \Big) \non \\
&&+\frac{\k^4}{16}  \Big( 
\la  v^2 {\bar v} \ra
-\la v  \ra \la  v {\bar v} \ra
-\hf \la  v^2 \ra \la  {\bar v}  \ra
+ \hf \la v  \ra^2 \la {\bar v } \ra  
~+~ {\rm c.c.}
\Big) \non \\
&&- \frac{\k^6}{64} \Big( 
\la v^2 {\bar v }^2 \ra
+\hf \la v {\bar v } v {\bar v} \ra
-\Big[ \la v  \ra
\la v {\bar v}^2 \ra
- {1 \over 4}  \la v \ra^2
\la  {\bar v}^2 \ra ~+~{\rm c.c.} \Big] 
\non \\
&& \qquad  \qquad 
+\la v  \ra \la {\bar v} \ra
\la v  {\bar v} \ra
- \hf \la v  {\bar v} \ra^2
-{1\over 4} \la v^2 \ra 
\la {\bar v}^2 \ra
-{1 \over 4} \la v  \ra^2
\la {\bar v} \ra^2
\Big)
\Big\} ~.
\label{AV2}
\eea
The fourth-order terms can be simplified
slightly:
\be 
{1 \over 4}  \int {\rm d}^4 x \, \Big(
\la v  +{\bar v} \ra^2 
- \la   (v  +{\bar v})^2 \ra \Big) 
=  \int {\rm d}^4 x \, \Big(
\la v  \ra \la {\bar v} \ra
- \la v  {\bar v} \ra \Big)~.
\ee
Regarding the eighth-order terms, 
the situation is more dramatic.  
Using the (easily verified) identities
\bea 
\la v^2 {\bar v }^2 \ra
&=& 
\Big( \la v \ra
\la v{\bar v }^2  \ra
- \hf \la v  \ra^2
\la {\bar v }^2 \ra + {\rm c.c.} \Big)
+\la v {\bar v } \ra
\Big( \la v {\bar v } \ra
- \la v  \ra \la {\bar v } \ra
\Big) 
~,  \non \\
2 \la v {\bar v } v {\bar v} \ra
&=&  
 \la v^2 \ra
\la {\bar v }^2 \ra 
-2 \la v{\bar v } \ra^2 
+ \Big( \la v  \ra^2
\la {\bar v }^2 \ra + {\rm c.c.} \Big)
+   \la  v \ra^2  
 \la  {\bar v} \ra^2 ~,
\eea
one can check that the eighth-order terms 
in (\ref{AV2}) completely cancel out! 
This result may seem strange, since 
the eighth-order terms in the AV action 
are, to the best of our knowledge, explicitly 
retained in all relevant publications,
starting with the classic papers
by Akulov and Volkov  \cite{VA,AV}
and continuing today,
e.g.  \cite{ST} (see, however
\cite{Lop} where it is demonstrated  
that the energy-momentum tensor 
for the AV model does not contain any
eighth-order terms). 
Therefore we will give another, 
purely algebraic and quite elementary,
proof. 
 
 The whole contribution from  
the eighth-order terms in 
the integrand in (\ref{AV2})
can be shown to be proportional to  
\be
 \ve_{abcd}\, \ve^{klmn} \,
{v_{k}}^{a} {v_{l}}^{b} 
{{\bar v}_{m}}\,\!\!^{c} {{\bar v}_{n}}\,\!\!^{d}
=  \l^{2} {\bar \l}^2 \, \ve_{abcd}\,\ve^{klmn} \, 
\left(\pa_{k} {\bar \l} \, {\tilde \s}^{ab} \, \pa_{l} {\bar \l}\right)
\, \left( \pa_{m} \l \, \s^{cd} \,\pa_{n} \l \right)~.
\ee
Using the well-known property of the 
sigma-matrices, 
\be 
\hf \,\ve_{abcd} \, \s^{cd} = - {\rm i} \,\s_{ab}~,
\quad 
\hf \,\ve_{abcd} \, \tilde{\s}^{cd} = {\rm i} \,\tilde{\s}_{ab}~,
\quad \longrightarrow \quad 
(\s^{ab})_\a{}^\b \,(\tilde{\s}_{ab})^\ad{}_\bd =0~,
\ee  
we see that 
the whole contribution under consideration vanishes.
As a result,  the AV action takes the form 
(\ref{AV3}).

\section{Old minimal supergravity: alternative realization}
\setcounter{equation}{0}
Here we consider an alternative realization of the 
model for old minimal supergravity  (\ref{omsg2}) 
that is obtained by making use of a variant 
superfield representation \cite{GS}
of the form
\be
\S^3 = -{1 \over 4} ({\bar \cD}^2 -4R) \,P~,
\qquad {\bar P} =P~, 
\ee
with $P$ an unconstrained real scalar superfield. 
It follows from (\ref{s-weyl-S}) that the super-Weyl 
transformation of $P$ is 
\be
P ~ \to ~  {\rm e}^{- \s -{\bar \s} } \,P~,
\label{s-weyl-P}
\ee
compare with (\ref{s-weyl-L}).
This implies that, for any real 
function $\cF(x)$ and  constant parameter $g$,  
the following action 
\bea
S &=& 
\intss\,E^{-1}\, P\,
\cF \Big( { {\bar \S} \, \S  \over P } \Big)
+ \Big\{ g \intss \, \ERc W^2 ~+~ {\rm c.c.} \Big\} ~,
\label{mass} \\
W_\a&=& -{1 \over 4} ({\bar \cD}^2 -4R )\cD_\a \, \ln P
\non
\eea
is super-Weyl invariant. 
This action turns out to describe 
supergravity provided $g=0$
and $\cF(x)$ is a linear function, 
$\cF(x) = -3x +\m$. Then we get
\bea
S &=& - 3
\intss\,E^{-1}\, {\bar \S} \, \S 
+\m \intss\,E^{-1}\, P \non \\
&=& -3
\intss\,E^{-1}\,  {\bar \S} \, \S  
+{\m \over 2} \, \Big\{  \intss \, \ERc  \,
\S^3 ~+~ {\rm c.c.} \Big\}~. 
\label{omsg3}
\eea
Here the second term on the right is a
supersymmetric cosmological term. 

In the family of actions (\ref{mass}), only 
the supergravity action (\ref{omsg3})
is invariant  under gauge transformations 
of the form
\be 
\d P = L~, \qquad 
({\bar \cD}^2 - 4R ) \, L = 0 
\ee 
that  leave $\S$ invariant.
More general models (\ref{mass}), which 
are generated by  a nonlinear function $\cF (x)$
and which involve the naked prepotential $P$, 
can be thought of as ``massive extensions''
of old minimal supergravity 
(compare with the unique
``massive extension''
of new minimal supergravity introduced
in \cite{K2}).

\end{appendix}

\end{document}